\RequirePackage{fix-cm}
\documentclass[smallextended]{svjour3}       
\smartqed  
\usepackage{graphicx}
\usepackage{amssymb,amsfonts,amsmath}
\begin{document}

\title{No alignment of cattle along geomagnetic field lines found}

\author{J. Hert         \and
        L. Jelinek         \and
        L. Pekarek         \and
        A. Pavlicek
}

\institute{J. Hert \at
              Department of Anatomy, Faculty of Medicine, Charles University, 305 06 Plzen, Czech Republic \\
              \email{jiri.hert@tiscali.cz}           
           \and
           L. Jelinek \at
              Department of Electromagnetic Field, Faculty of Electrical Engineering, Czech Technical University in Prague, 166 27 Prague, Czech Republic \\
              Tel.: +420 224 355 891\\
              Fax: +420 233 339 958\\
              \email{l\underline{ }jelinek@us.es}           
					\and
           L. Pekarek \at
              National Reference Laboratory for Electromagnetic Field, National Institute of Public Health, 100 42 Prague, Czech Republic \\
              \email{pekarek@szu.cz}           
					\and
           A. Pavlicek \at
              Department of System Analysis, Faculty of Informatics and Statistics, University of Economics, Prague, Czech Republic \\
              \email{antonin.pavlicek@vse.cz}           
}

\date{Received: date / Accepted: date}
%
\maketitle
\begin{abstract}
This paper presents a study of the body orientation of domestic cattle on free pastures in several European states, based on Google satellite photographs. In sum, 232 herds with 3412 individuals were evaluated. Two independent groups participated in our study and came to the same conclusion that, in contradiction to the recent findings of other researchers, no alignment of the animals and of their herds along geomagnetic field lines could be found. Several possible reasons for this discrepancy should be taken into account: poor quality of Google satellite photographs, difficulties in determining the body axis, selection of herds or animals within herds, lack of blinding in the evaluation, possible subconscious bias, and, most importantly, high sensitivity of the calculated main directions of the Rayleigh vectors to some kind of bias or to some overlooked or ignored confounder. This factor could easily have led to an unsubstantiated positive conclusion about the existence of magnetoreception.

\end{abstract}
\section*{Introduction}
\label{intro}
The possibility that many animals, including large mammals, possess the ability to sense a magnetic field and use it for some benefit has been widely discussed in the scientific community, see for example \cite{Johnsen2005} and the references therein. This is understandable, as the question of sensitivity of living creatures to a magnetic field is also interesting from the point of view of human health protection \cite{Schuz2008}. Reports of the discovery of magnetic sensing in large mammals such as cattle and deer, published in \cite{Begall2008} and supported by observations of the behavior of cattle near power lines \cite{Burda2009} also point to possible health risks for humans.

Evidence of magnetic sensing of migrating and nonmigrating animals has been based mostly on observations of animal behavior \cite{Wiltschko,Dennis2007}, as the responsible biophysical mechanism has remained in the state of hypotheses \cite{Johnsen2005}. The three proposed mechanisms, one based on the presence of ferromagnetic or ferrimagnetic microcrystals in the body tissue \cite{Kirschvink2001}, another based on electric currents induced in the body of an animal moving in the Earth's magnetic field \cite{Kalmijn1978}, and the third based on chemical reactions connected with the influence of a magnetic field on free radicals and their recombination \cite{Ritz2000}, are potential candidates, but there is as yet no final proof for any of them.

To obtain a sufficient number of observations to allow statistically significant conclusions, a large amount of data - usually considerably dispersed and sometimes contradictory \cite{Wiltschko} - needs to be collected. This may be expensive and time-consuming. The mentioned problem of time and expense was successfully resolved in \cite{Begall2008}: To find whether ruminants (mostly domestic cows) possess a magnetic sense, Begall et al. evaluated the angles between the local direction of the magnetic declination and the direction of the body axes of more than eight thousand animals seen on satellite photographs in several parts of the globe. Along the data obtained in this way, Begall et al. also studied the angles of almost 3000 body axes of wild ruminants (red deer and roe deer), obtained by direct field glass observations and by examining the prints left behind in show by resting or sleeping wild animals. The evaluation of the data obtained in that way indicated that the investigated animals, if not disturbed, orient their body axes parallel to the lines of magnetic declination. The authors also supported the idea of the ability of ruminants to perceive a magnetic field by observations of cows in the vicinity of power lines \cite{Burda2009}, where the animals allegedly lost their magnetic sense, and the direction of their body axes became chaotic. 

The findings published in \cite{Begall2008} are unexpected and call for replication. In this paper, we present the results of an evaluation of 3412 cows seen in satellite pictures. We compare them with the results published in \cite{Begall2008}, and suggest an explanation for the difference between our results and the results published in \cite{Begall2008}. In order to obtain results as robust as possible, we divided the evaluation into three parts. First, we evaluated the orientation of the herds allowing a direct comparison with \cite{Begall2008}. Second, we evaluated the body axes of individual cows, as in our opinion magnetoreception is a property of an individual and not of the herd. Third, we evaluated the orientation of the heads, as it is possible that, even if no unidirectionality were found in the axial data, all cows could be oriented with their heads to the northern half of the circle.

\section*{Material and methods}
\label{sec:1}
Two groups (Hert, Jelinek, Pekarek -- 1636 animals, and Pavlicek -- 1776 animals) independently collected the angles of 3412 body axes and 589 body vectors of cows from 232 herds in France, Great Britain, Germany, the Netherlands, Switzerland and the Czech Republic, using satellite screenshots from Google Earth and a computer program for angle evaluation (for angle acquisition we used the vector graphic environment for marking the angles. The marked angles were then exported in text format as values), see ``Online Resource 1'' for the GPS coordinates and the angles. 

In accordance with the way chosen in \cite{Begall2008}, only animals on pastures located in horizontal areas sufficiently apart from communications and other disturbing arrangements were included in the data. No data used for evaluation overlap. Only data with recognized head positions (589 animals) was used twice, first as circular data with vector directions between $0^\circ$ and  $360^\circ$, and, for the axial statistics, transformed to axial data and added to other data of animals with head direction not safely recognizable. In order to account for the uncertainty of measurement of the angles of the animals, a computer--generated jitter in the range of $\pm 4^\circ$ (estimated error) was applied to the measured data.

Full quantitative information about the resulting distribution of the observed angles was visualized via circular histograms, both for the vector data and for the axial data. To keep a balance between the precise representation and the unavoidable imprecision in data acquisition, we chose the histogram step to be $10^\circ$.

Apart from the visual information contained in the circular histograms, the uniformity Z--test (in our case coinciding with the Rayleigh test \cite{Mardia}) defined as

\begin{equation}
{\text{Z}}_m^2  = \frac{2}
{N}\left( {\left( {\sum\limits_i {\cos \left( {m\varphi _i } \right)} } \right)^2  + \left( {\sum\limits_i {\sin \left( {m\varphi _i } \right)} } \right)^2 } \right)
\label{eq1}
\end{equation}
was also evaluated on the measured sample \cite{Mardia}, with $\varphi _i \in \left\langle {0,2\pi /m} \right\rangle $ and $N$ as the total number of samples. In \eqref{eq1}, $m = 1$ and $m = 2$ were used for the vector data and for the axial data, respectively. For uniformly (isotropically) distributed angles $\varphi _i $, the quantity ${\text{Z}}_m^2 $ has a chi-square distribution with two degrees of freedom. The graphical data is also accompanied by numerical values of the mean sample direction $\varphi_{\text{mean}}$ \cite{Mardia} and of the circular sectors $\alpha_{\text{north}}$, $\alpha_{\text{mean}}$, centered around north or mean direction, respectively, containing 50 \% of the sample. Note that uniform distribution should have $\alpha_{\text{north}} = \alpha_{\text{mean}} = 180^\circ/m$.

\section*{Results}
\label{sec:2}
Before we move to the main part of the paper, which deals with the data statistics, we will first clarify how the evaluation will be made. The procedure consists of two steps. In the first step we will present the numerical values of the Z-statistics and the mean direction. The second step consists of a visual inspection of the circular histogram. If, in the first step, the Z-statistics does not exceed the significance level, the evaluation is terminated with the conclusion that the sample cannot be distinguished from a uniform distribution. The mean direction in this case has no significance. If, on the other hand, the Z-statistics exceeds the significance level, uniformity of the sample is rejected. Then, we need a second step, in which the kind of non-uniformity (single modal around a mean angle, multimodal, etc.) is analyzed. 

In order to clarify this way of evaluation, we present following example: Assume an artificial sample of 1000 body axes coming from a uniform distribution on the interval $(10^\circ,170^\circ)$. For this sample, the Z-statistics will reach values of approximately 50, which greatly exceeds the significance level. In addition, this sample will show a mean angle of approximately $90^\circ$. If we were to take only the numerical values of $Z$ and of the mean angle, we would draw the completely wrong conclusion that the sample is significantly north-south aligned. That such judgment is wrong is immediately seen from the plot of the probability density, which reveals that the high value of $Z$ is only connected with ``holes'' in the east-west direction. In this case, we can only say that the individuals avoid orienting themselves in a sharply defined cone around the east-west direction. If, on the other hand, the 1000 body axes are selected from a Gaussian distribution with mean value $90^\circ$ and $\sigma = 10^\circ$, the Z-statistics will also exceed the significance level (now by a value around 2000), and the calculated mean direction will also read approximately $90^\circ$. A visual inspection of the corresponding probability density plot will, however, clearly show that in this case the non-uniformity is caused by a sharp north-south alignment.

Statistics emerging from the real collected data will now be presented.

First, we evaluated the orientation of the herds, in order to obtain results directly comparable to \cite{Begall2008}. For the evaluation, we chose only those herds that contained more than four cows, leading to 110 herds in Group I and 110 herds in Group II. The distribution of the mean axes of the herds is shown in Fig. 1, and leads to ${\text{Z}}_2^2  = \left[0.2,5.5\right]$, $\varphi_{\text{mean}}=\left[148^\circ,82^\circ \right]$, $\alpha_{\text{north}}=\left[90^\circ,74^\circ\right]$, $\alpha_{\text{mean}}=\left[87^\circ,75^\circ\right]$ for group I and for group II, respectively. Comparing the values of ${\text{Z}}_2^2$ with the 0.95-quantile of a chi-square distribution with two degrees of freedom, which is equal to 6.0, shows that at this significance level data uniformity cannot be rejected. This conclusion agrees with the visual inspection of Fig. 1a,b, where no preferred direction is observed. 

\begin{figure*}[hbt]
\begin{center}
\centerline{\includegraphics[width=0.75\textwidth]{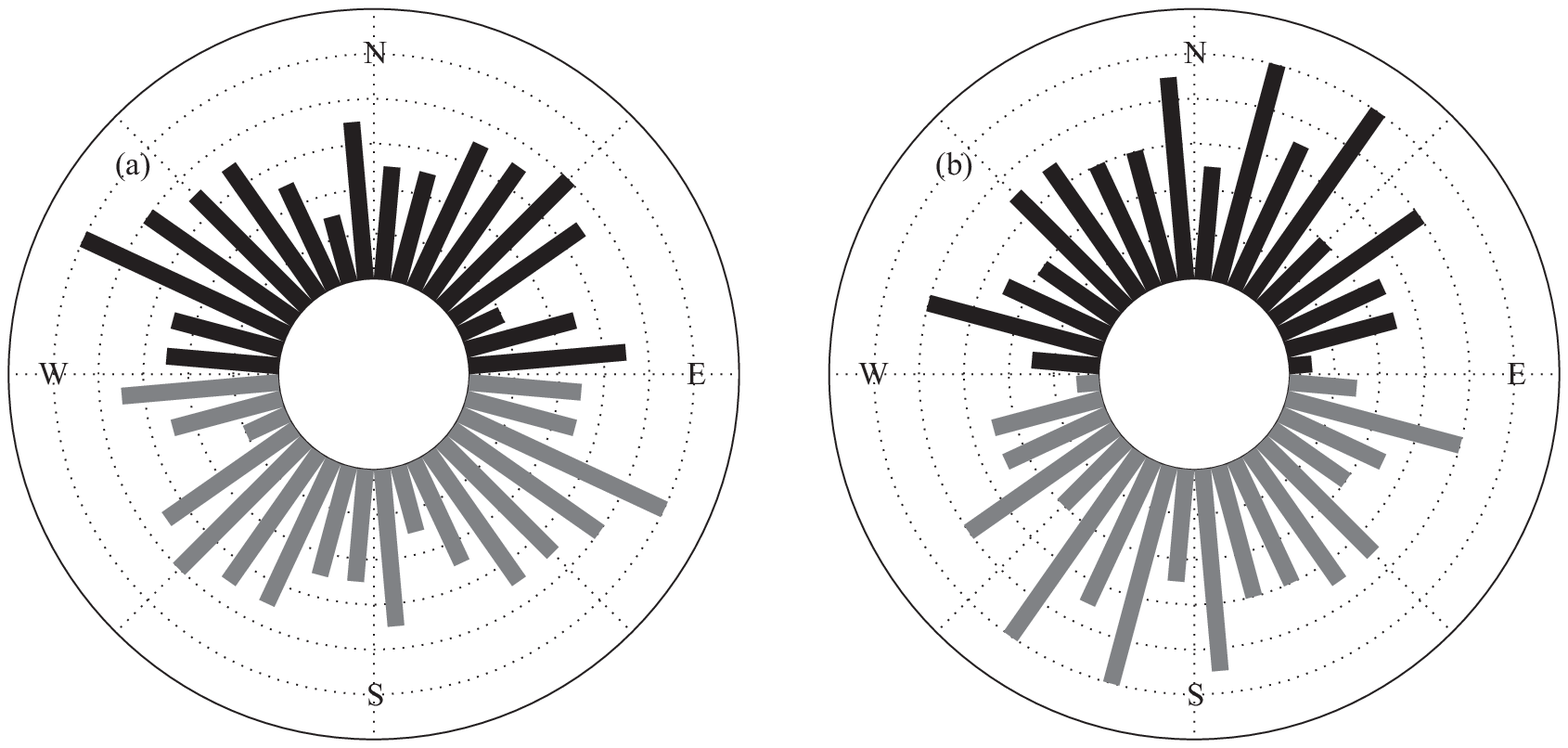}}
\caption{Axial data showing the orientation of the mean vectors of the herds. Panels (a,b) represent data found for group I and group II, respectively. Each rectangular beam in the figure covers an angular section of $10^\circ$ and its length represents the number of herds. Each line of the circular grid represents 2 herds. The bottom gray part of the figures represents the centro-symmetric image.}\label{fig1}
\end{center}
\end{figure*}

In the second step, we evaluated the body axes of individual cows. The results are shown in Fig. 2, and are characterized by ${\text{Z}}_2^2  = \left[1.2,14.9\right]$, $\varphi_{\text{mean}}=\left[148^\circ,91^\circ \right]$, $\alpha_{\text{north}}=\left[91^\circ,85^\circ \right]$, $\alpha_{\text{mean}}=\left[90^\circ,85^\circ \right]$ for Fig. 2a (group I) and Fig. 2b (group II), respectively. A comparison of the values of ${\text{Z}}_2^2$ with the 0.95-quantile of the chi-square distribution with two degrees of freedom shows that at this level of significance uniformity cannot be rejected in the case of Group I. Group II, however, has to be denoted as nonuniform. The nonuniformity is apparent from Fig. 2b, which shows that the distribution is very close to the hypothetical case of uniform distribution with east-west ``holes'' that has been discussed above and cannot be interpreted as a north-south alignment. 

\begin{figure*}[hbt]
\begin{center}
\centerline{\includegraphics[width=0.75\textwidth]{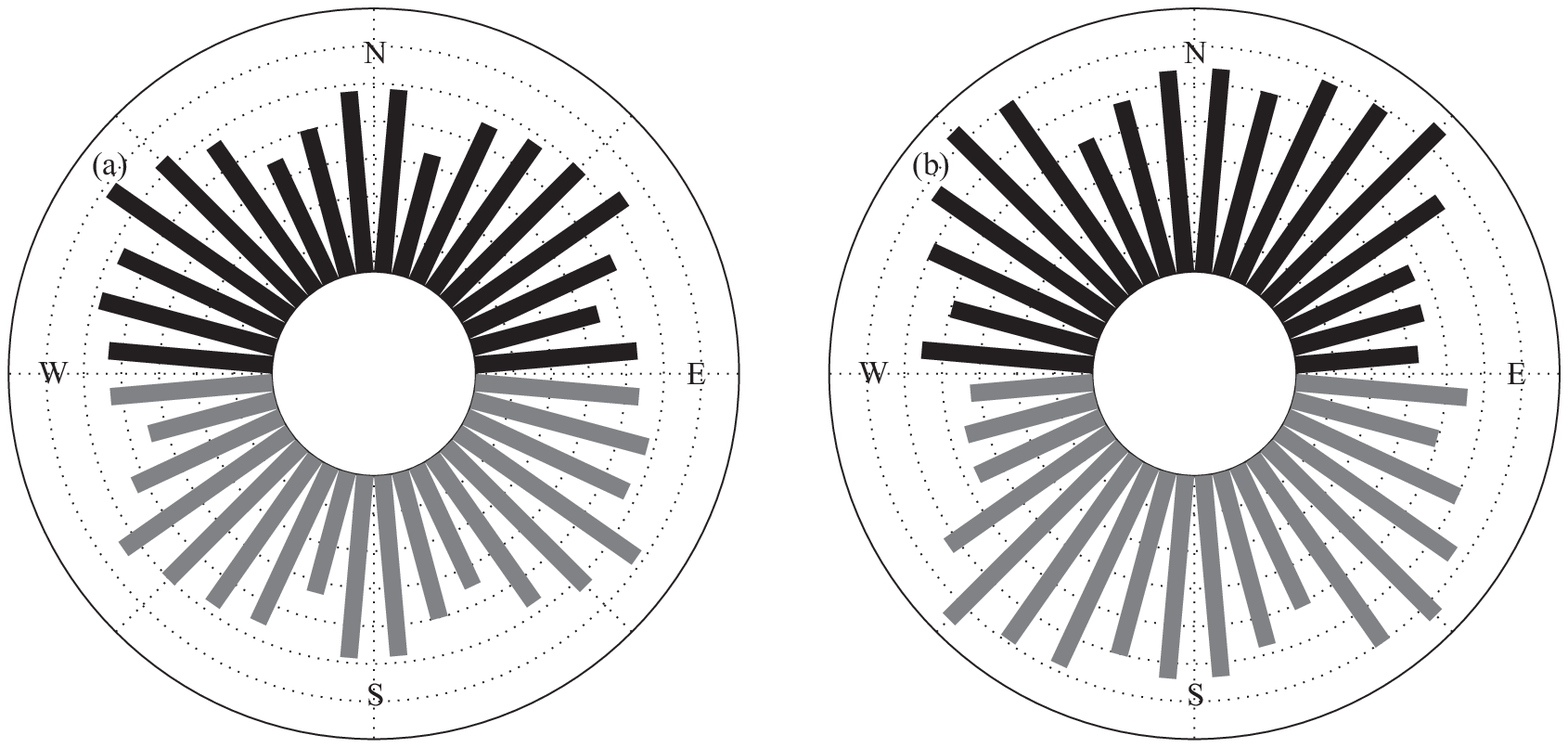}}
\caption{Axial data showing the orientation of the mean vectors of the cows. Panels (a,b) represent data found by group I and group II, respectively. Each rectangular beam in the figure covers an angular section of $10^\circ$ and its length represents the number of cows. Each line of the circular grid represents 20 cows. The bottom gray parts represent the centro-symmetric image.}\label{fig2}
\end{center}
\end{figure*}

In the last step, we evaluated the head orientation of individual cows. In order to get an idea about the directional distribution of the heads, Fig. 3 shows the collection of 20 probability density diagrams belonging to randomly selected herds of cows with a recognized head position. With an average number of 18 cows in one herd, the diagrams distinctly reveal that the vectors of the bodies are not distributed randomly, being clustered in a small number of directions (mostly two). Only the diagram in the row 2, column 2 displays a pattern with two opposite directions aimed to the north and to the south. The other diagrams show less tidy patterns and no apparent preference for the north or for the south, but clustering of the body angles still remains apparent. Looking at the diagrams, we can guess that the probability density distribution of the full sample will not be uniform, but no preferred direction can be expected.

\begin{figure*}[hbt]
\begin{center}
\centerline{\includegraphics[width=0.75\textwidth]{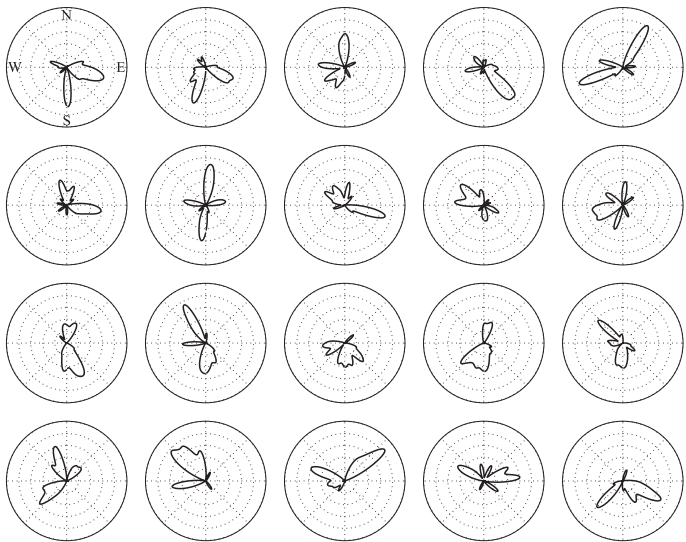}}
\caption{Probability density function drawn for 20 herds randomly selected from 40 herds with well resolved positions of heads and with a similar number of cows (361 animals in total). For better visibility, the density functions were approximated by a Fourier series with 18 harmonics, which leads to angular resolution of $10^\circ$.}\label{fig3}
\end{center}
\end{figure*}

\begin{figure*}[hbt]
\begin{center}
\centerline{\includegraphics[width=0.375\textwidth]{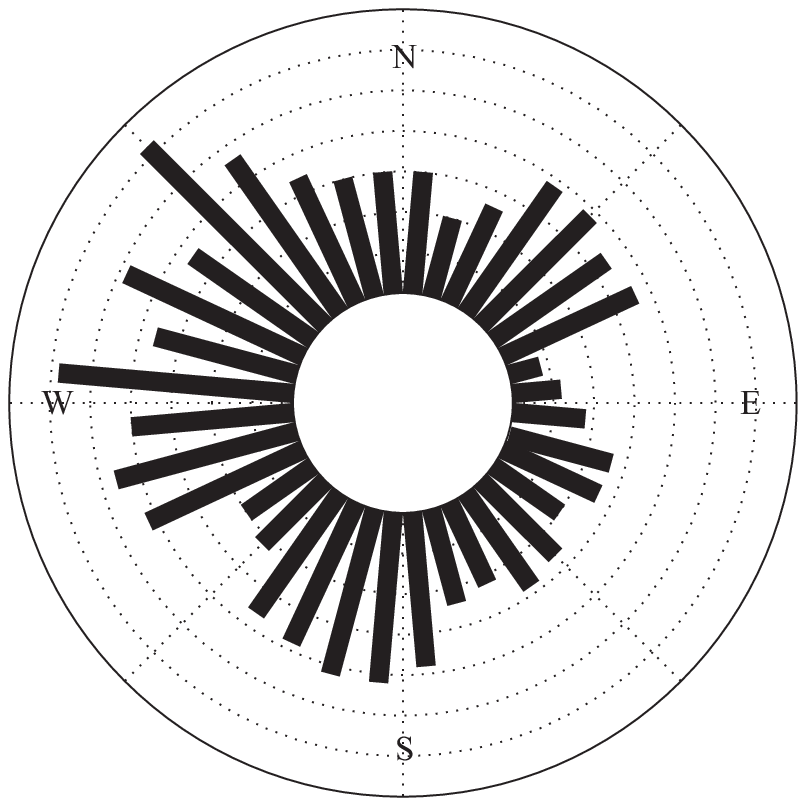}}
\caption{Vector data sample based on herds with a recognized head position. Each rectangular beam in the figure covers an angular section of $10^\circ$ and its length represents the number of cows. Each line of the circular grid represents 5 cows.}\label{fig4}
\end{center}
\end{figure*}

A graphical representation of all vector samples (cows with recognized head position) fused together is shown in Fig. 4, and leads to ${\text{Z}}_1^2  = 27.9$, $\varphi_{\text{mean}}=168^\circ$, $\alpha_{\text{north}}=170^\circ$, $\alpha_{\text{mean}}=150^\circ$. Comparing the values of ${\text{Z}}_1^2$ given above with the 0.95-quantile of the chi-square distribution with two degrees of freedom leads to rejection of uniformity of the measured data. This conclusion agrees with a visual inspection of Fig. 4, which shows several maxima and minima. The figure is however far from being sharply north aligned, which is also supported by the numerical values of $\alpha_{\text{north}}$, $\alpha_{\text{mean}}$.

Summarizing our results acquired from the data obtained from the satellite photographs, we must unequivocally state that no tendency of the animals to prefer the north-south direction could be extracted from an evaluation of the pictures. The orientation of the herds, the body axes (position of head ignored) or the body vectors (position of head recognized), was either isotropic (uniform distribution), or multimodal.

\section*{Discussion}
\label{sec:3}
In our study, performed by two independent groups of authors in three different ways, no north-south alignment of domestic cattle on free pastures was found. Our result is thus essentially different from the result obtained in \cite{Begall2008}. Certainly, not all our results can be compared with the results published in \cite{Begall2008}, where the orientation of the heads was not included in the evaluated data, while we undertook no observation of the behavior of wild ruminants or of cows under power lines \cite{Burda2009}. However, the weight of the visual observation of the body directions of wild deer or of the direction estimated from prints in snow is due to inherent inaccuracy and possible bias certainly lower than the measurement of body angles on satellite pictures, which can be carried out in a ``double blind'' way (the screen shots of cows can be evaluated by several uninformed persons in unknown position after rotation). In addition, if no north-south alignment was found with cows on a free pasture, then the idea of disruption of the magnetic sense by a weak alternating magnetic field near power lines lacks sense. 

Hence the crucial problem is to explain the fundamental difference between the two outcomes of evaluations of axial data obtained from satellite pictures. The difference in the number of animals investigated cannot serve as an explanation, since the size of our sample is even larger than the European sample from \cite{Begall2008}, which clearly showed north-south alignment. Many other factors must be taken into consideration: poor quality of Google photographs, difficulties in determining the body axis, different samples of herds, selection of herds or animals within the herds, omitted blinding of the evaluation, possible subconscious bias, incorrect or erroneous statistics, high sensitivity of the mean vectors to some kind of bias, or some overlooked or ignored confounder. 

It is difficult to detect the main reason, but one essential difference is apparent between the two studies. The major part of our work has used cows as basic entities, unlike \cite{Begall2008}, where the chosen basic entities were herds. In other words, in our case the calculated mean vector of each herd would have its own amplitude and direction, while in the case of \cite{Begall2008} all herd vectors were unitary. The method of unitary vectors, however, completely ignores the number of cows in each herd, which can lead to significant error when there are big differences between herd sizes (a herd with a single cow is given the same significance as a herd with 100 cows). The method of unitary vectors also completely ignores the statistical dispersion within a herd (a herd with 99 uniformly distributed cows and 1 cow oriented to the north is given the same significance as a herd with 100 north oriented cows). This error can be further enhanced by data selection, as not all herds found in the satellite maps were used for the evaluation. The restrictions claimed in \cite{Begall2008} for data that can be accepted for further processing are certainly justified, but the method with individual animals chosen as basic units is certainly more immune against unintentional bias. Unitary weight for each herd is also difficult to understand. Indeed, there are no clues in the literature for thinking that the presumed magnetoreception should be a common property of a herd, and not an individual property of each cow. Certainly one or more animals standing nearby tend to assume the same body orientation, but most animals during pasture are dispersed over the field.

The most probable reason, why prevalence of the north-south alignment of cattle was found in \cite{Begall2008}, is, in our opinion, an inadequate selection of herds and/or individual cows. Fortunately, the method for evaluating satellite photographs used in \cite{Begall2008} for investigating the behavior of large-sized animals offers a simple, effective and freely accessible way to collect the corresponding data. Hence, there is an easy way to replicate the investigations on possible magnetic alignment of cattle, and to ascertain whether or not the magnetic alignment of grazing or resting cattle is reality. We hope that our study will stimulate such replications.

\begin{acknowledgements}
This work has been supported by National Institute of Public Health and the Czech Grant Agency (project no. 102/09/0314). We also would like to thank Radek Theier, Filip Truhlar, Jan Prochazka, Filip Kroupa, and Michael Hapala for their assistance in data acquisition.
\end{acknowledgements}


\begin{thebibliography}{30}

\bibitem{Johnsen2005}
Johnsen S and Lohmann KJ (2005) The physics and neurobiology of magnetoreception.  NAT REV NEUROSCI 6: 703–-712

\bibitem{Schuz2008}
Schuz J, Ahlbom A (2008) Exposure to electromagnetic fields and the risk of childhood leukemia: a review. RADIAT PROT DOSIM 132: 202–-211

\bibitem{Begall2008}
Begall S, Cerveny J, Neef J, Vojtech O, Burda H (2008) Magnetic alignment in grazing and resting cattle and deer. P NATL ACAD SCI USA 105: 13451–-13455

\bibitem{Burda2009}
Burda H, Begall S, Cerveny J, Neef J, Nemec P (2009) Extremely low-frequency electromagnetic fields disrupt magnetic alignment of ruminants. P NATL ACAD SCI USA 106: 5708--5713

\bibitem{Wiltschko}
Wiltschko R, Wiltschko W (1995) Magnetic Orientation in Animals. Springer, Berlin

\bibitem{Dennis2007}
Dennis TE, Rayner MJ, Walker MM (2007) Evidence that pigeons orient to geomagnetic intensity during homing. P ROY SOC B-BIOL SCI 274: 1153–-1158

\bibitem{Kirschvink2001}
Kirschvink JL, Walker MM and Diebel CE (2001) Magnetite-based magnetoreception. CURR OPIN NEUROBIOL 11: 462--467

\bibitem{Kalmijn1978} 
Kalmijn AJ (1978) Animal Migration, Navigation, and Homing. Springer, Berlin

\bibitem{Ritz2000}
Ritz T, Adem S, Schulten K (2000) A Model for Photoreceptor-Based Magnetoreception in Birds. BIOPHYS J 78:707--718

\bibitem{Mardia}
Mardia KV, Jupp PE (2000) Directional Statistics. John Wiley and Sons, Chichester

\end{thebibliography}
\end{document}